\documentclass[12pt]{article}   

\usepackage{amsmath,amssymb,amsfonts,latexsym}
\usepackage{graphicx,psfrag}
\usepackage{float}
\usepackage{geometry}
\usepackage{booktabs}
\usepackage{caption}
\usepackage[table]{xcolor}
\usepackage[most]{tcolorbox}
\usepackage{xcolor}
\usepackage{indentfirst}        
\tcbuselibrary{listingsutf8}

\newtcbox{\alphatag}{on line,
  colback=gray!15!white, colframe=gray!50!black,
  boxrule=0pt, arc=4pt, boxsep=0pt, left=4pt, right=4pt, top=2pt, bottom=2pt,
  fontupper=\ttfamily\footnotesize, nobeforeafter }

\title{Deep Learning for Short-Term Equity Trend Forecasting: \\
A Behavior-Driven Multi-Factor Approach}
\author{Yuqi Luan\thanks{Master of Science UZH ETH in Quantitative Finance}}

\begin{document}
\maketitle

\begin{abstract}

This study applies machine learning to short-term stock return prediction using behavioral alpha factors. We evaluate MLP, CNN, and SVM models on 40 technical signals, and find that MLP achieves the best performance in IC and portfolio returns. Results highlight the value of behavior-driven modeling for active investment strategies.
\end{abstract}
\vspace{7ex}

\section{Introduction}

In recent years, machine learning (ML) has emerged as a powerful tool in financial modeling, particularly in the realm of active portfolio management. Unlike traditional factor models that rely on linear specifications and limited economic assumptions, ML techniques allow for flexible modeling of complex, nonlinear relationships between market signals and asset returns. This flexibility is particularly valuable in short-horizon trading environments, where investor behavior, market microstructure effects, and rapidly changing sentiment play a significant role.

Short-term market movements are often driven not only by fundamentals but also by behavioral patterns such as momentum chasing, panic selling, bottom-fishing, and volume-price divergence. Capturing these subtle, transient behaviors requires a more dynamic approach to alpha generation. While conventional factor models struggle to adapt to the non-stationarity and noise of high-frequency markets, machine learning models—especially those based on deep neural networks—have demonstrated promising performance in uncovering latent structures in noisy financial data.
\vspace{2ex}
This paper proposes a behavior-driven, scenario-specific modeling framework for short-horizon investment signals. We construct a rich set of features combining 40 formulaic alpha factors with four customized behavioral indicators aimed at detecting three key market patterns: volume-price divergence, bottom reversal, and momentum-driven price chasing. By separating these behavioral regimes and designing tailored model structures for each (classification, regression, or hybrid), we aim to improve the precision and interpretability of signal-based forecasting.
\vspace{2ex}
To validate our approach, we implement and compare multiple machine learning architectures—namely multilayer perceptrons (MLP), convolutional neural networks (CNN), and support vector machines (SVM)—to assess predictive power. Model outputs are evaluated using return-based metrics such as Information Coefficient (IC), top-K performance, and heatmap visualizations. Our results demonstrate that behavior-specific DL models can extract valuable alpha from short-term market data, supporting their application in active portfolio strategies.

\vspace{5ex}

\section{Objective}

The objective of this study is to develop a deep learning-based trading algorithm for short-horizon equity investments. Rather than predicting market indices directly, we construct behavior-driven signals on individual stocks, aiming to identify and exploit transient patterns rooted in investor psychology—such as momentum chasing, panic selling, volume-price divergence, and bottom reversals. These behaviors are modeled as either classification or regression problems, depending on whether the task is to detect the occurrence of a pattern or estimate its future impact on returns.
\\ \\
Trading fundamentally relies on prediction. Every decision to buy or sell implies an expectation of future price movements. In active portfolio management, this predictive component is central to alpha generation. The more precisely a model can capture near-term behavioral shifts, the more actionable and profitable its signals become.
\\ \\
For example, in momentum-chasing scenarios, a stock may exhibit consecutive short-term gains, rising RSI values, and expanding trading volumes—suggesting herd-driven buying behavior. This is not the result of a single indicator but of multiple features acting jointly and often nonlinearly. If the price rises without volume support, it may be a false breakout; but when combined with volume spikes and overbought signals, the likelihood of sustained momentum increases.
\\ \\
Such compound patterns are difficult to capture using traditional models. Therefore, we adopt multilayer perceptrons (MLPs) as our primary predictive architecture. MLPs are well-suited to learning nonlinear relationships and interactions between features, automatically assigning weights to combinations of factors. This allows the model to represent complex behavioral structures that arise from the co-movement of technical and volume-based signals.
\\ \\
By designing scenario-specific models and applying them to a diverse set of factor combinations, our approach aims to generate precise, interpretable short-term predictions that can support active trading decisions and enhance index-relative performance.

\section{Network Architectures}
\begin{figure}[htbp]
    \centering
    \includegraphics[width=0.65\linewidth]{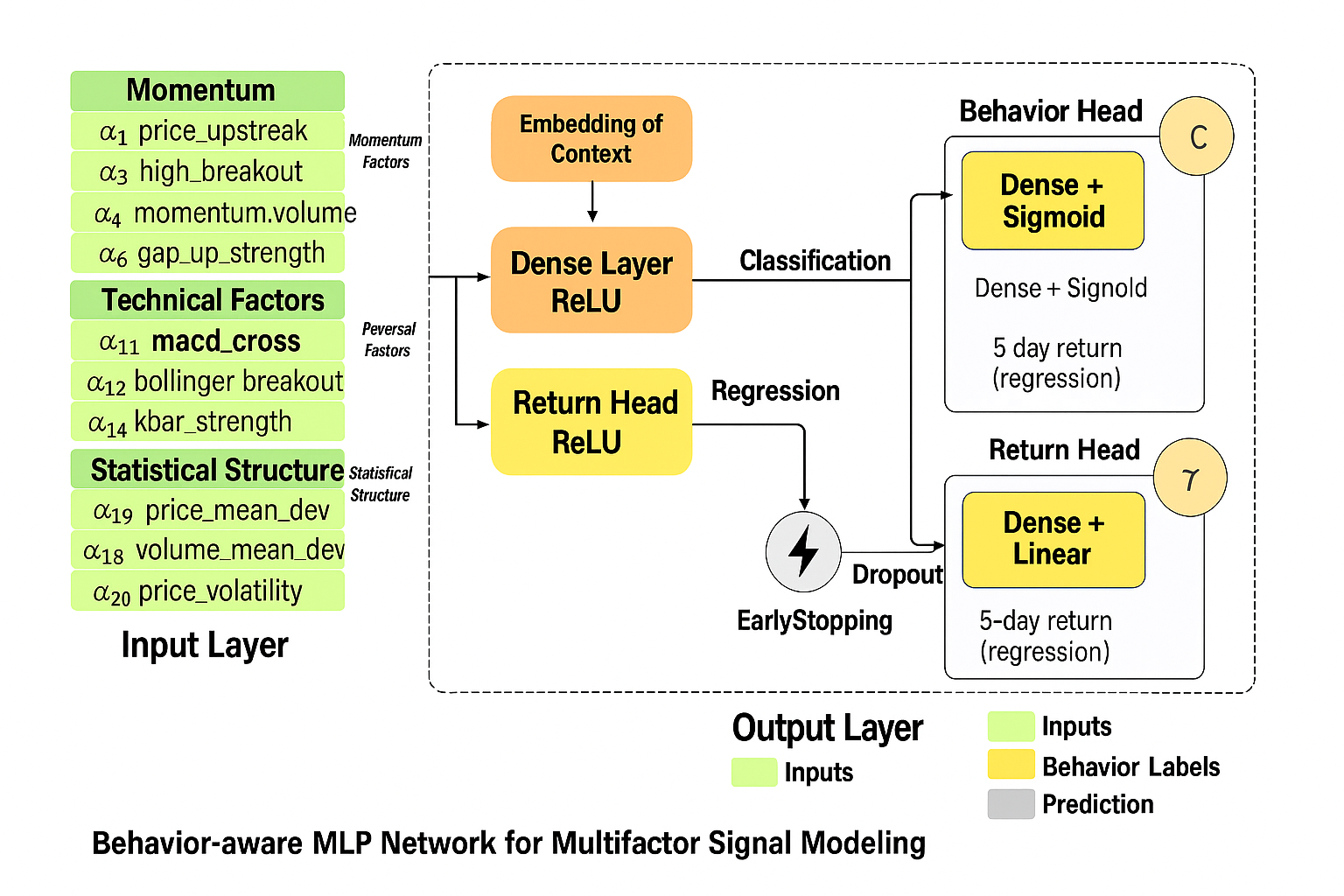}
    \caption{\quad Behavior-aware MLP Network for Multifactor Signal Modeling}
    \label{fig:mlp_structure}
\end{figure}

\subsection{Multilayer Perceptron (MLP) Architecture}

We define a multilayer perceptron (MLP) as a deep feedforward neural network composed of $L$ layers, each parameterized by a weight matrix $\mathbf{W}^{(l)}$ and bias vector $\mathbf{b}^{(l)}$. Let the input feature vector be $\mathbf{x} \in \mathbb{R}^{d}$, representing the set of $d$ alpha factors and behavioral indicators. The network computes a mapping $f: \mathbb{R}^{d} \rightarrow \mathbb{R}$ (for regression) or $f: \mathbb{R}^{d} \rightarrow [0, 1]$ (for binary classification), using a sequence of nonlinear transformations:

\begin{align}
\mathbf{h}^{(0)} &= \mathbf{x}, \\
\mathbf{h}^{(l)} &= \sigma\left( \mathbf{W}^{(l)} \mathbf{h}^{(l-1)} + \mathbf{b}^{(l)} \right), \quad \text{for } l = 1, \dots, L-1, \\
\hat{y} &= \phi\left( \mathbf{W}^{(L)} \mathbf{h}^{(L-1)} + \mathbf{b}^{(L)} \right),
\end{align}

where:
\begin{itemize}
  \item $\sigma(\cdot)$ is an element-wise non-linear activation function, typically ReLU: $\sigma(z) = \max(0, z)$,
  \item $\phi(\cdot)$ is the output activation function: identity for regression, sigmoid for binary classification, or softmax for multi-class classification.
\end{itemize}

The learnable parameters of the model are $\Theta = \{ \mathbf{W}^{(l)}, \mathbf{b}^{(l)} \}_{l=1}^L$. The network is trained on a dataset $\{(\mathbf{x}_i, y_i)\}_{i=1}^{N}$ to minimize a loss function $\mathcal{L}(\Theta)$.

For regression (e.g., return prediction), we use the mean squared error (MSE):
\begin{equation}
\mathcal{L}_{\text{reg}} = \frac{1}{N} \sum_{i=1}^{N} \left( y_i - \hat{y}_i \right)^2.
\end{equation}

For binary classification (e.g., predicting behavior labels), we use binary cross-entropy loss:
\begin{equation}
\mathcal{L}_{\text{clf}} = -\frac{1}{N} \sum_{i=1}^{N} \left[ y_i \log(\hat{y}_i) + (1 - y_i)\log(1 - \hat{y}_i) \right].
\end{equation}

Model training is performed using mini-batch stochastic gradient descent (SGD) or Adam optimizer, with gradients computed by backpropagation. To improve generalization, regularization techniques such as dropout, $L_2$-norm penalty, and early stopping are applied.
Start with deep feed--forward nets (``dense layers''), then describe AE, RNN and CNN in some detail. Explain what you can do for regularization. 
\vspace{2ex}
\subsection{Recurrent/ Convolutional Networks}

Convolutional Neural Networks (CNNs) are widely used to extract local patterns from structured data. In financial applications, we treat each sample as a matrix $X \in \mathbb{R}^{T \times F}$, where $T$ is the lookback window and $F$ is the number of alpha factors.

A convolutional layer applies a filter $W \in \mathbb{R}^{k_t \times k_f}$ across $X$ to compute local features:

\[
Z_{i,j} = \sum_{m=0}^{k_t - 1} \sum_{n=0}^{k_f - 1} X_{i+m, j+n} \cdot W_{m,n}
\]

The result passes through a non-linear activation, typically ReLU:
\[
\text{ReLU}(z) = \max(0, z)
\]

After optional pooling, the features are flattened and fed into fully connected layers. The final prediction is:
\[
\hat{y} = W^\top h + b
\]

The model is trained by minimizing a loss function (e.g., MSE for regression), using gradient descent and backpropagation.
\vspace{2ex}

\vspace{5ex}

\section{Experiments}

\subsection{Choice of signals}
\textbf{Alpha Factor Design and Behavioral Structure Classification}

To effectively model nonlinear market behaviors in short-term trading cycles, this study constructs 40 technical alpha factors grounded in behavioral finance theory. These factors are derived from high-frequency price and volume data, incorporating elements such as open price, close price, high-low range, volume dynamics, and standard technical indicators including MACD, RSI, and Bollinger Bands. Unlike traditional indicators, which often fail to reflect the complexity of real-world market regimes, our alpha factors are built through nonlinear combinations of multiple behavioral signals. Specifically, we apply rank transformations, conditional logic, and multiplicative interactions (e.g., RSI × MACD, price deviation × volume spike) to model structure-specific patterns. This approach captures three essential market microstructure signals: panic-driven sell-offs, momentum-driven rallies, and oversold reversals. The resulting alpha features embed multidimensional investor behavior into interpretable signals that are well-suited for machine learning applications such as multilayer perceptrons (MLPs).
\vspace{2em}

\begin{itemize}

    \item \textbf{Bottom Reversal}: This structure typically occurs after a significant price decline during a consolidation or rebound phase. It is characterized by oversold recovery, price rebound, and mild volume expansion. Representative factors include 
    \begin{alphatag}
    {alpha\_rsi\_bounce\_strength}
    \end{alphatag} and 
    \begin{alphatag}{alpha\_macd\_cross\_strength}\end{alphatag}, which reflect potential reversal signals from RSI rebounds and MACD golden cross events.

    \item \textbf{Volume-Price Divergence}: When price movement diverges from volume trends, it may indicate weakening momentum or an upcoming reversal. Representative factors such as 
    \begin{alphatag}
    {alpha\_volspike\_times\_body}
    \end{alphatag} and 
    \begin{alphatag}
    {alpha\_macd\_times\_lowdev}
    \end{alphatag} are used to identify abnormal behaviors like “price up with volume down” or “price down with volume up”.

    \item \textbf{Momentum and Herding}: This structure features strong trending behavior and herd-following trading patterns, often observed in one-sided market rallies or crashes. Key factors such as 
    \begin{alphatag}
    {alpha\_momentum\_5d\_min\_rank}
    \end{alphatag} and 
    \begin{alphatag}{alpha\_macd\_rsi\_product}\end{alphatag} capture enhanced momentum and trend-following behavior among traders.
\end{itemize}

The factor construction follows these principles:

\begin{itemize}
    \item \textbf{Structure-first identification}: Market conditions such as price location, shape, and momentum are identified first, then combined with volume or volatility to confirm signals.

    \item \textbf{Behavioral cross-features}: Factors are built by combining multiple technical indicators (e.g., MACD $\times$ RSI, Volume $\times$ Volatility) to better reflect investor psychology.

    \item \textbf{Normalization and Ranking}: Many factors use \texttt{rank}-based transformations to reduce sensitivity to extreme values and enhance robustness in model training.
\end{itemize}

\vspace{3em}

\subsection{Choice of network type and parameters}
\vspace{2em}
\vspace{1em}

\subsubsection{Network Architecture: Dual-task MLP Model}

To evaluate the effectiveness of behavioral alpha factors in short-term return prediction, we construct a simple yet robust deep learning model --- a \textbf{dual-task multilayer perceptron (MLP)}. The model simultaneously performs two tasks: \textit{regression} to predict 5-day forward return, and \textit{classification} to identify whether the asset will rise, leveraging a shared representation to jointly learn both objectives.

\paragraph{Architecture Design}
\begin{itemize}
    \item \textbf{Input Layer}: 40-dimensional input vector, consisting of over 20 behavioral alpha factors and several short-term technical indicators.
    \item \textbf{Hidden Layer 1}: 64 neurons with ReLU activation; Dropout rate = 0.1.
    \item \textbf{Hidden Layer 2}: 32 neurons with ReLU activation; Dropout rate = 0.1.
    \item \textbf{Output Heads}:
    \begin{itemize}
        \item \textbf{Regression Head}: Predicts continuous future 5-day return.
        \item \textbf{Classification Head}: Predicts the binary outcome label\_up, i.e.,` whether return is positive, with a Sigmoid activation.
    \end{itemize}
\end{itemize}
\vspace{2ex}

\paragraph{Training Setup}
All weights are initialized using Xavier uniform initialization, and biases are set to zero. The model is trained using the Adam optimizer with:
\[
\text{learning rate} = 5 \times 10^{-4}, \quad \text{weight decay} = 1 \times 10^{-3}.
\]
The loss function is a weighted combination of regression loss (MSE) and binary classification loss (BCE):
\[
\mathcal{L}_{\text{total}} = \mathcal{L}_{\text{reg}} + 0.5 \times \mathcal{L}_{\text{cls}}.
\]

We apply \textbf{gradient clipping} (max norm = 0.5) to ensure numerical stability. The learning rate is dynamically reduced using a plateau scheduler (factor = 0.7, patience = 5). To prevent overfitting, early stopping is adopted with a patience of 15 epochs based on validation RMSE.

\vspace{1em}
\paragraph{Preprocessing}
All input features are standardized using cross-sectional z-score normalization \textit{per day} to remove scale differences across assets. The target variable \texttt{future\_return\_5} is clipped to the [5\%, 95\%] quantile range to suppress the impact of extreme outliers. We use a time-based split: data before 2023 is used for training, and data from 2023 onward is reserved for validation.

\paragraph{Model Motivation}
This dual-task architecture enhances the model's ability to understand both trend magnitude and directional movement. It reflects the behavioral finance hypothesis that short-term market movements are driven by a blend of \textit{emotion}, \textit{momentum}, and \textit{reversal} effects. The shared feature representation allows both heads to benefit from a more expressive latent structure, leading to better generalization and stronger signal extraction.

\vspace{0.5em}
In summary, this dual-head MLP provides a lightweight yet powerful benchmark for short-cycle behavioral modeling in financial time series.
After training, the model is used to generate a continuous predictive signal referred to as \textit{deep\_alpha}, which is evaluated using IC, IR, and top-K backtesting. Daily deep\_alpha values are compared against realized 5-day forward returns to compute the Spearman correlation (IC). We also visualize the signal distribution, IC histogram, and return scatter plots for interpretation.

Furthermore, a long-short backtest is conducted by selecting the top-5 and bottom-5 stocks by \textit{deep\_alpha} score each day. The cumulative return curves for top, bottom, and long-short portfolios are plotted and analyzed.

\vspace{5ex}

\subsubsection{Choice of Network Type and Parameters (CNN)}

To capture short-term temporal behavior and reduce feature redundancy, we implement a lightweight one-dimensional convolutional neural network (CNN) as a comparative baseline. Unlike the MLP, the CNN uses local receptive fields to learn sequential interactions across the alpha factor space.

The CNN architecture is structured as follows:
\begin{itemize}
    \item \textbf{Input Layer}: a $40 \times 1$ feature tensor;
    \item \textbf{Convolutional Layer}: one 1D convolution with kernel size 3 and 8 output channels, followed by ReLU activation;
    \item \textbf{Fully Connected Layers}:
    \begin{itemize}
        \item one hidden dense layer with 32 units and ReLU activation;
        \item one output layer producing the 5-day forward return prediction.
    \end{itemize}
\end{itemize}

We initialize all weights using Xavier uniform initialization and set all biases to zero. The model is optimized using the Adam optimizer with a learning rate of $5 \times 10^{-4}$ and L2 regularization (weight decay $= 1 \times 10^{-3}$). Gradient clipping with a maximum norm of 0.5 is applied to ensure numerical stability. We use early stopping with a patience of 15 epochs and a learning rate scheduler that reduces the learning rate by a factor of 0.7 when validation loss plateaus for 5 consecutive epochs.

The preprocessing pipeline is consistent with the MLP model: all input features are standardized using cross-sectional z-scores per day, and target returns are clipped to the 5th and 95th percentiles to mitigate outliers. A temporal split is adopted, using data before 2023 for training and data from 2023 onward for validation.
\vspace{2ex}
\subsubsection{SVM Model}

As a baseline model, we adopt a linear Support Vector Regression (SVM) to fit the relationship between the 40-dimensional factor vector and the 5-day forward return. The model is implemented using the default configuration of \texttt{scikit-learn}'s \texttt{SVR(kernel='linear')} with standard $L_2$ regularization. It contains no nonlinear transformations or deep structures and directly uses all training samples without parameter tuning.

To ensure consistency with the deep learning models, the input features undergo the same preprocessing steps:
\begin{itemize}
    \item Daily cross-sectional z-score standardization;
    \item Clipping of future 5-day returns to the 5\%--95\% percentile range.
\end{itemize}
\vspace{2ex}

\vspace{2ex}
\subsubsection{Evaluation Metric}

To evaluate signal quality and portfolio relevance, we use five key metrics:

\paragraph{ Information Coefficient (IC):}

The IC measures cross-sectional rank correlation between model scores $\hat{y}_t$ and future 5-day returns $y_t$:

\begin{equation}
\mathrm{IC}_t = \rho(\hat{y}_t, y_t) = \frac{1}{N} \cdot \frac{(\hat{y}_t - \bar{\hat{y}}_t)^\top (y_t - \bar{y}_t)}{\mathrm{std}(\hat{y}_t) \cdot \mathrm{std}(y_t)}
\end{equation}

where $N$ is the number of stocks on day $t$. A higher IC indicates better ranking ability of the model.

\paragraph{ IC Information Ratio (ICIR):}

\begin{equation}
\mathrm{ICIR} = \frac{\mathbb{E}[\mathrm{IC}_t]}{\mathrm{std}(\mathrm{IC}_t)}
\end{equation}

ICIR measures the consistency of IC values over time and reflects signal robustness.

\paragraph{Information Ratio (IR)}

\begin{equation}
\mathrm{IR} = \frac{\mathbb{E}[\mathrm{IC}_t]}{\mathrm{std}(\mathrm{IC}_t)}
\end{equation}

The Information Ratio (IR) measures the consistency of the predictive signal over time. It is defined as the ratio of the mean daily Information Coefficient (IC) to its standard deviation. A higher IR indicates not only stronger average ranking power but also greater temporal stability in the signal quality.

\paragraph{Sharpe Ratio:}

We construct daily long-short portfolios using top and bottom model scores, and compute the Sharpe ratio:

\begin{equation}
S_p = \frac{r_p - r_f}{\sigma_p}
\end{equation}

Here, $r_p$ is the average daily return of the portfolio, $r_f$ is the risk-free rate (set to 0), and $\sigma_p$ is its return standard deviation.

This metric captures the stability of predictive signals from a practitioner’s perspective.
\subsection{Results}
Comparison of IC

\begin{figure}[H]
  \centering
  \includegraphics[width=0.5\textwidth]{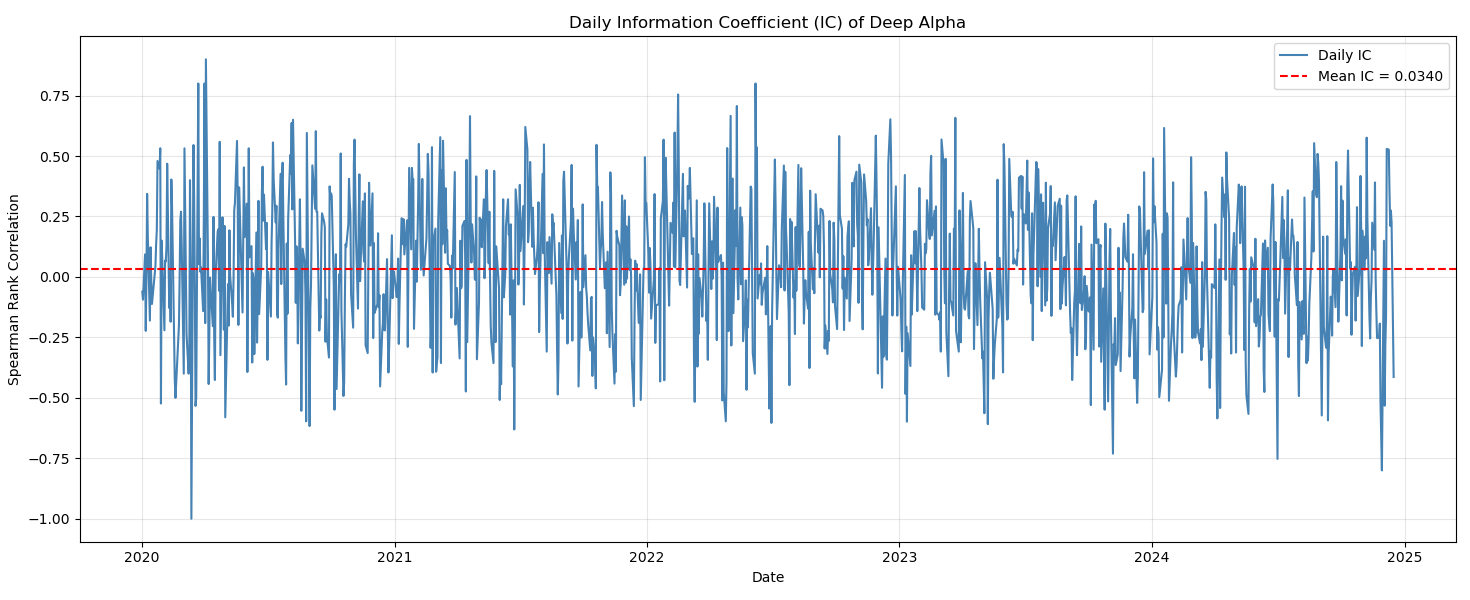}
  \caption{IC Information Ratio ( Double Task MLP Model)}
  \label{fig:cnn}
\end{figure}
\vspace{1em}  

\begin{figure}[H]
  \centering
  \includegraphics[width=0.5\textwidth]{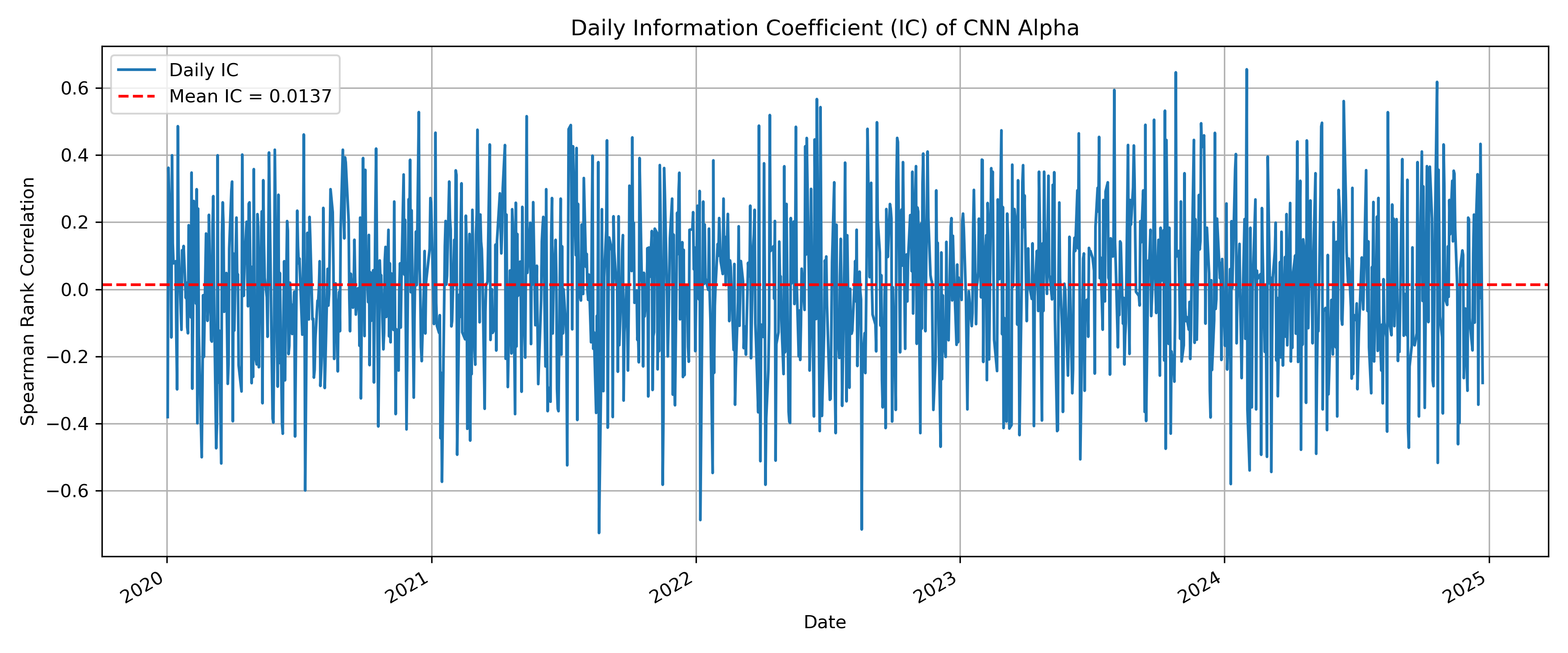}
  \caption{IC Information Ratio ( Double Task CNN Model)}
  \label{fig:cnn}
\end{figure}
\vspace{1em}  

\begin{figure}[H]
  \centering
  \includegraphics[width=0.5\textwidth]{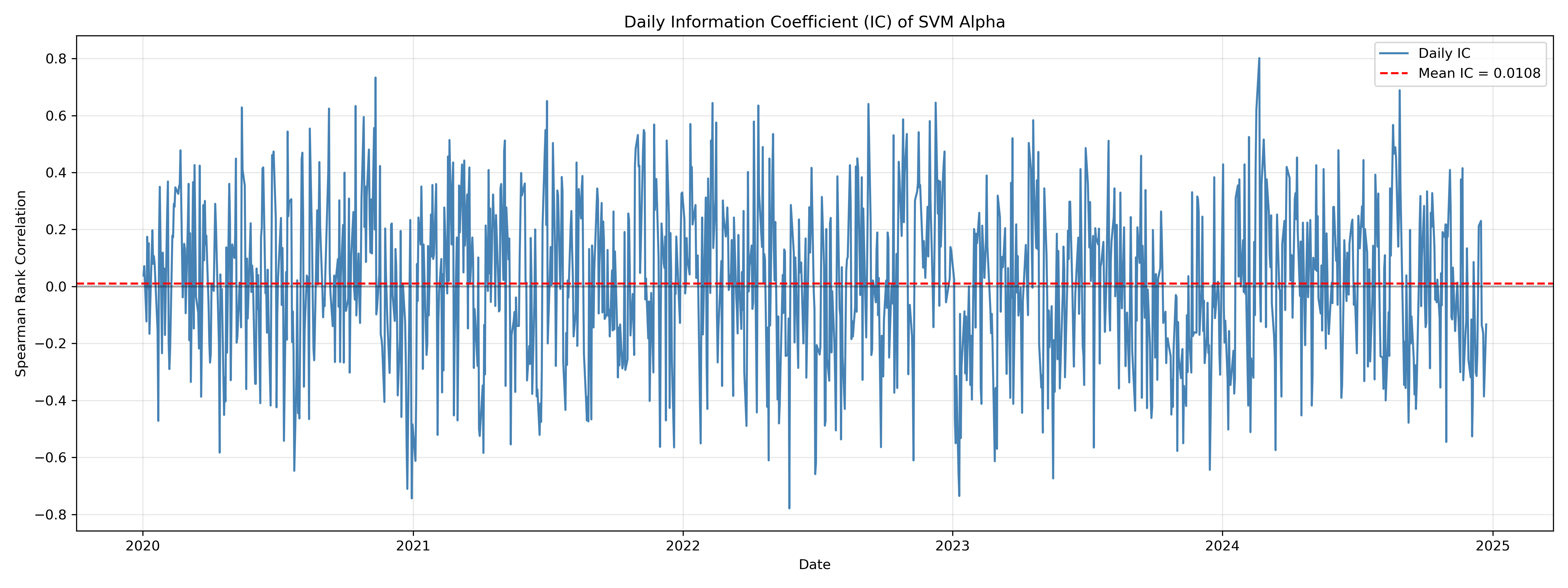}
  \caption{IC Information Ratio ( SVM Model)}
  \label{fig:cnn}
\end{figure}
\vspace{1em}  

\begin{table}[htbp]
\centering
\caption{Comparison of IC, IR, and ICIR Metrics for Three Models}
\renewcommand{\arraystretch}{1.2}
\begin{tabular}{@{}lcccc@{}}
\toprule
\textbf{Model} & \textbf{IC $\uparrow$} & \textbf{IC Std $\downarrow$} & \textbf{IR $\uparrow$} & \textbf{ICIR $\uparrow$} \\
\midrule
SVM             & 0.0108 & 0.2084 & 0.0518 & 0.0518 \\
CNN             & 0.0137 & 0.1998 & 0.0686 & 0.0686 \\
\textbf{MLP (Dual-task)} & \textbf{0.0340} & \textbf{0.2097} & \textbf{0.1621} & \textbf{0.1621} \\
\bottomrule
\end{tabular}
\label{tab:icir_comparison}
\end{table}
\vspace{2em}
\subsubsection*{P\textbf{erformance Analysis based on IC, IC Std, and IR}}

To evaluate cross-sectional predictive performance, we analyze three key metrics: the Information Coefficient (IC), its standard deviation (IC Std), and the Information Ratio (IR). The IC measures the Spearman rank correlation between model scores and future returns on each day, indicating the signal’s ranking power. IC Std captures the variability of this ranking power, and IR — defined as the ratio of IC to IC Std — balances signal strength and stability. A higher IR indicates a model that produces consistently strong signals across time, making it a robust indicator of overall performance.

As shown in Figure 2 and Table I, the \textbf{MLP dual-task model} significantly outperforms both CNN and SVM in terms of IC magnitude and consistency. It achieves the highest average daily IC of \textbf{0.0340}, with a relatively low IC Std of \textbf{0.2097}, resulting in an IR of \textbf{0.1621}, the strongest among the three models. This suggests that the MLP model not only produces accurate signals but also maintains stability over time. The \textbf{CNN model} delivers moderate performance with a mean IC of \textbf{0.0137}, but its higher IC Std of \textbf{0.1998} reduces its IR to \textbf{0.0686}, reflecting increased volatility in its predictive signals. In contrast, the \textbf{SVM model} performs the weakest, with the lowest IC (\textbf{0.0108}) and a similar IC Std (\textbf{0.2084}), yielding the lowest IR (\textbf{0.0518}). This indicates that SVM provides both weak and unstable alpha signals in a cross-sectional context.

\vspace{2em}

\begin{figure}[H]
  \centering
  \includegraphics[width=0.5\textwidth]{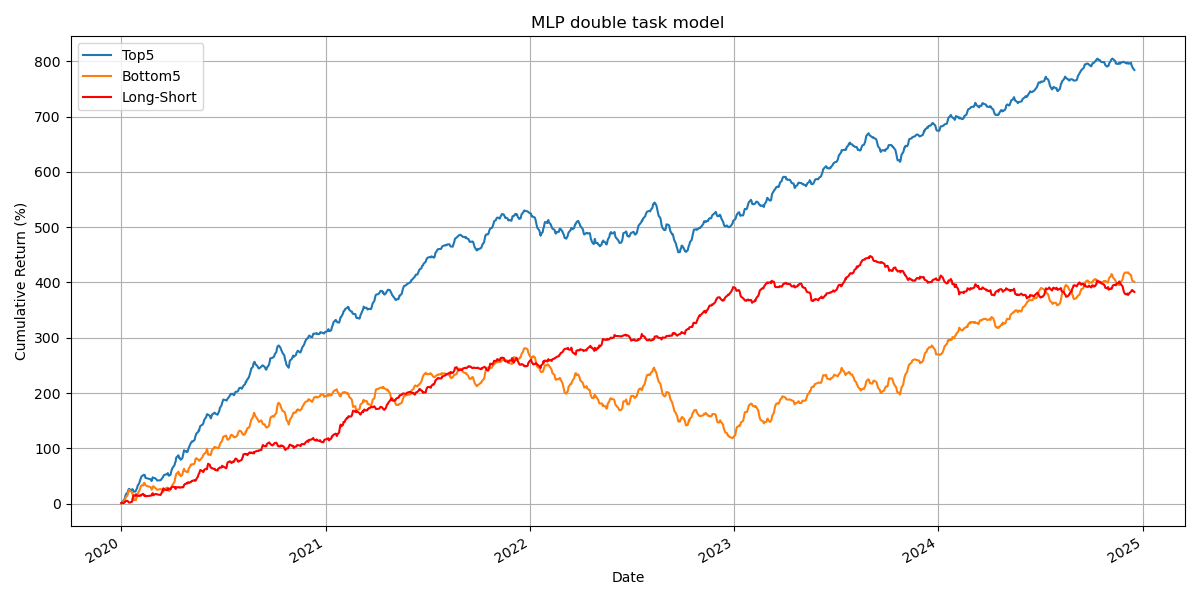}
  \caption{Cumulative Return Curve (Dual-task MLP Model)}
  \label{fig:cnn}
\end{figure}
\vspace{1em}  
\begin{figure}[H]
  \centering
  \includegraphics[width=0.5\textwidth]{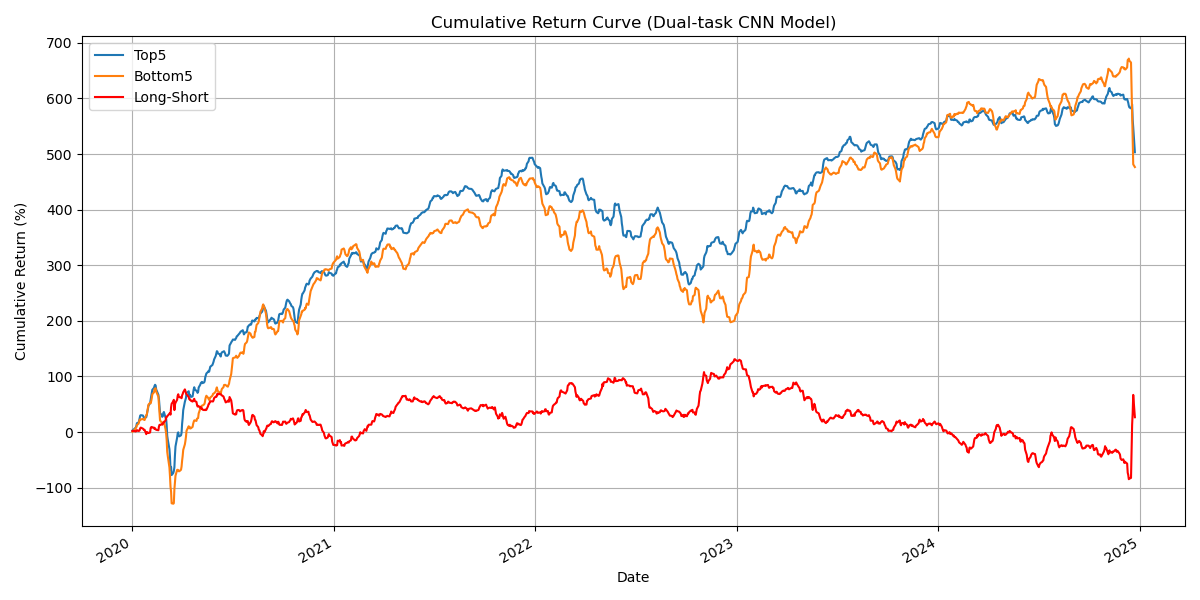}
  \caption{Cumulative Return Curve (Dual-task CNN Model)}
  \label{fig:cnn}
\end{figure}
\vspace{1em}  
\begin{figure}[H]
  \centering
  \includegraphics[width=0.5\textwidth]{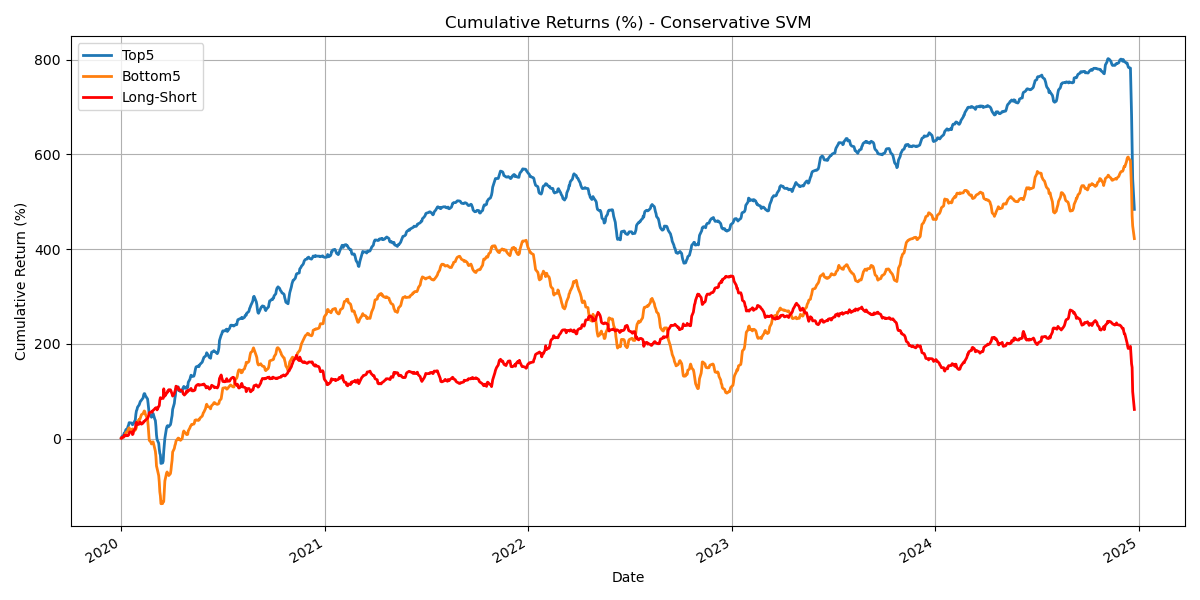}
  \caption{Cumulative Return Curve (SVM Model)}
  \label{fig:cnn}
\end{figure}
\vspace{1em}  

As shown in the cumulative return curves, the dual-task MLP model outperforms both the CNN and SVM baselines across the entire test period. The Top 5 cumulative return in 5 years for MLP model reaches above 800\%.  The long-short strategy based on MLP exhibits a smoother and more robust upward trend, with minimal drawdowns and consistent gains. In contrast, the CNN model yields more volatile returns, and the SVM model performs the worst, remaining mostly flat with weak signal differentiation.
\vspace{1em}

The superior performance of the MLP model can be attributed to its dual-task architecture, which jointly learns both regression (future return magnitude) and classification (up/down direction). This setup allows the model not only to rank stocks effectively but also to identify directional momentum, enhancing its ability to filter out noisy signals. As a result, the MLP long-short return consistently outperforms the bottom-group return, indicating its ability to capture alpha beyond simple bottom-avoidance.
\vspace{1.5em}

Interestingly, during the sharp market downturn in January 2025—coinciding with a significant drop in the Nasdaq due to AI sector risks—the CNN and SVM strategies suffered severe drawdowns, while the MLP model showed more resilience. This suggests that the MLP’s directional learning provides stronger robustness under extreme market conditions, likely due to better signal filtering and more informed position allocation.

\vspace{1em}
\begin{table}[ht]
\centering
\caption{Comparison of Annualized Sharpe Ratios Across Models}
\begin{tabular}{lccc}
\hline
\textbf{Model} & \textbf{Annualized Return} & \textbf{Annualized Volatility} & \textbf{Sharpe Ratio} \\
\hline
MLP  & 0.5349 & 0.3321 & \textbf{1.6075} \\
CNN  & 0.4570 & 0.3980 & 1.1487 \\
SVM & 0.2500 & 0.3244 & 0.7709 \\
\hline
\end{tabular}
\label{tab:sharpe_comparison}
\end{table}

Sharpe Ratio Comparison: MLP vs. CNN vs. SVM
To evaluate the effectiveness of different modeling approaches in extracting trading signals from behavior-driven factors, we compare the annualized Sharpe Ratios of three models: a multi-task MLP, a dual-task CNN, and a classical linear SVM. The results show that the MLP achieves the highest Sharpe Ratio (1.6075), followed by the CNN (1.1487), while the SVM lags significantly behind (0.7709). This performance gap highlights the advantages of deep learning models in capturing complex, nonlinear patterns inherent in market behaviors such as momentum, reversals, and volume-price divergences.
\vspace{1em}

The MLP’s superior performance can be attributed to its ability to simultaneously learn regression and classification objectives, enhancing signal robustness. The CNN benefits from spatial structuring of input factors but may suffer from oversmoothing in certain temporal regimes. In contrast, the SVM’s underperformance stems from its limited capacity to model nonlinear relationships, lack of feature interaction learning, and sensitivity to noise and outliers, making it less suitable for high-dimensional financial data. Overall, the results demonstrate that neural networks, particularly those employing multi-task learning, offer superior return efficiency when modeling behavior-aware alpha factors.
\vspace{5em}

\begin{figure}[H]
  \centering
  \includegraphics[width=0.7\textwidth ]{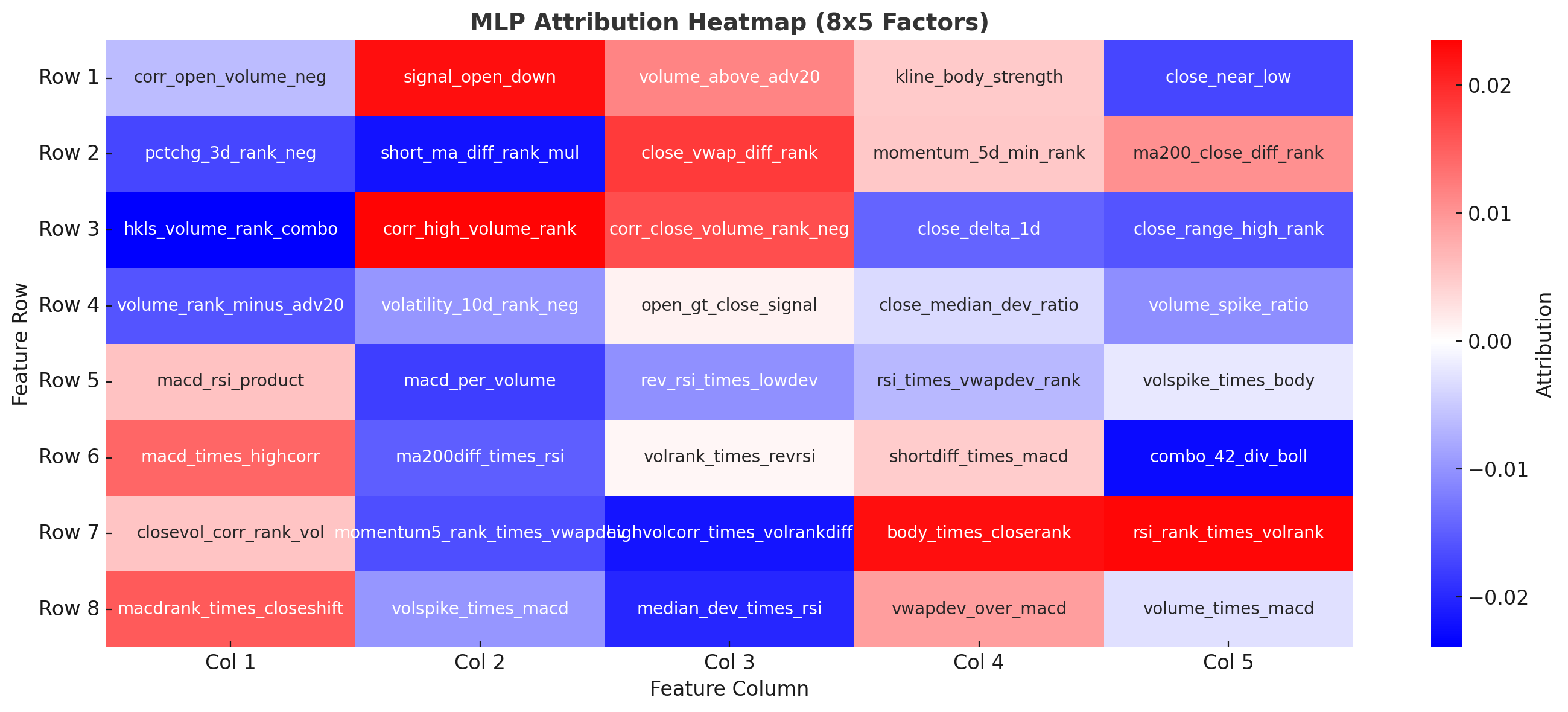}
  \caption{2-task MLP Attribution Heatmap)}
  \label{fig:cnn}
\end{figure}
\vspace{1em}  %
To interpret the role of individual alpha factors in the dual-task MLP model, we compute SHAP values and generate an attribution heatmap. Each cell color in the heatmap indicates the SHAP attribution of a factor on the model's predicted return: red implies a positive contribution, while blue suggests a negative one. The 40 alpha factors are manually arranged into an 8$\times$5 matrix for clearer visualization.

From the heatmap, we observe that factors such as \begin{alphatag}{macd\_rsi\_product}\end{alphatag}, \begin{alphatag}{momentum5\_rank}\end{alphatag}, and \begin{alphatag}{volume\_above\_adv20}\end{alphatag}
 exhibit the highest positive attribution scores, suggesting their dominant role in capturing short-term upward trends. Conversely, factors like \begin{alphatag}{volatility\_10d}\end{alphatag}, \begin{alphatag}{macd\_times\_lowdev}\end{alphatag}, and \begin{alphatag}{vwapdev\_over\_macd}\end{alphatag} consistently demonstrate negative contributions, potentially linked to trend exhaustion signals or noise sensitivity.

Interestingly, this attribution pattern aligns with our predefined factor structure. Specifically:
\begin{itemize}
    \item \textbf{Momentum \& Herding} factors are dominant in the positively attributed region, indicating their relevance in capturing trend-following behaviors.
    \item \textbf{Volume-Price Divergence} factors appear more frequently in negatively attributed areas, consistent with their reversal detection nature.
    \item \textbf{Bottom Reversal} factors show a more neutral distribution, reflecting structural stability across market phases.
\end{itemize}

\section{Conclusion}
This study demonstrates that deep learning, particularly the dual-task MLP framework, offers substantial advantages in multi-factor stock selection by integrating short-cycle technical and behavioral signals into a unified predictive model. Compared with traditional linear and shallow machine learning approaches, deep networks provide superior capacity to capture nonlinear relationships and complex factor interactions, enabling the detection of behavioral market patterns such as bottom reversals, volume–price divergences, and momentum-driven herding. Empirical results confirm that the dual-task MLP consistently outperforms CNNs and SVMs in terms of Information Coefficient (IC), Information Ratio (IR), and portfolio backtests, indicating both higher predictive accuracy and greater robustness in alpha signal extraction. The architecture’s ability to jointly forecast return magnitudes and directional movements proves particularly effective in filtering noise and enhancing stability under varying market conditions.

Despite these achievements, the application of deep learning to multi-factor forecasting faces important challenges. Overfitting remains a concern, as performance depends heavily on data volume and regime stability, with abrupt structural shifts potentially undermining reliability. Interpretability also remains limited: while SHAP-based attributions provide some insight, the black-box nature of neural networks hinders intuitive understanding of factor contributions and investor behavior. Moreover, the current experimental design abstracts away from trading frictions, as transaction costs, liquidity constraints, and turnover effects are not fully embedded in the optimization process. This gap between predictive modeling and implementable portfolio strategies highlights the need for end-to-end approaches. Finally, heterogeneous performance across architectures—such as the weaker stability of CNNs—raises questions about generalization across markets and regimes.

Looking forward, several directions merit exploration. Structure-aware multitask learning could enrich predictive frameworks by incorporating behavioral event labels, allowing models to jointly recognize structural signals and forecast returns. Regime-aware architectures, leveraging gating or mixture-of-experts mechanisms informed by volatility, turnover, or breadth, may enhance adaptability under dynamic environments. Improving interpretability through monotonicity constraints or deep lattice networks would better align predictions with financial priors. End-to-end portfolio optimization, embedding costs and risk constraints directly into training, would bridge the gap between theoretical accuracy and practical tradability. Finally, robust cross-market and regime-switching validation is essential to ensure stability and resilience in real-world applications.

In conclusion, deep learning significantly advances the frontier of multi-factor alpha discovery, yet its long-term value depends on addressing challenges of interpretability, robustness, and implementation. Future research should focus on developing structure-aware, regime-adaptive, and trading-conscious frameworks that unify predictive accuracy with practical portfolio management.

\section{Appendix}
\begin{table}[H]
\centering
\scriptsize
\begin{tabular}{ll}
\toprule
\textbf{Factor Name} & \textbf{Formula} \\
\midrule
alpha\_signal\_open\_down & (Open \textless
 Open.shift(1)).astype(int) \\
alpha\_kline\_body\_strength & (Close - Open) / (High - Low + 0.001) \\
alpha\_close\_near\_low & (High - Close) / (High - Low + 0.001) \\
alpha\_short\_ma\_diff\_rank\_mul & rank(Close - ma5) * rank(Close - ma20) \\
alpha\_close\_vwap\_diff\_rank & rank(Close - vwap) \\
alpha\_momentum\_5d\_min\_rank & -1 * rank(Close - Close.shift(5)) \\
alpha\_ma200\_close\_diff\_rank & rank(ma200 - Close) \\
alpha\_close\_delta\_1d & -1 * Close.diff() \\
alpha\_close\_range\_high\_rank & rank((Close - Low) / (High - Low + 0.001)) \\
alpha\_volatility\_10d\_rank\_neg & -1 * rank(std(Close, 10)) \\
alpha\_close\_median\_dev\_ratio & (Close - (High+Low)/2) / (High - Low + 0.001) \\
alpha\_volume\_spike\_ratio & Volume / ma(Volume, 5) \\
alpha\_macd\_rsi\_product & macd\_diff * rsi\_14 \\
alpha\_macd\_per\_volume & macd\_diff / Volume \\
alpha\_rev\_rsi\_times\_lowdev & (100 - rsi\_14) * alpha\_close\_near\_low \\
alpha\_rsi\_times\_vwapdev\_rank & rsi\_14 * alpha\_close\_vwap\_diff\_rank \\
alpha\_volspike\_times\_body & alpha\_volume\_spike\_ratio * alpha\_kline\_body\_strength \\
alpha\_ma200diff\_times\_rsi & alpha\_ma200\_close\_diff\_rank * rsi\_14 \\
alpha\_volrank\_times\_revrsi & alpha\_volatility\_10d\_rank\_neg * (100 - rsi\_14) \\
alpha\_shortdiff\_times\_macd & alpha\_short\_ma\_diff\_rank\_mul * macd\_diff \\
alpha\_median\_dev\_times\_rsi & alpha\_close\_median\_dev\_ratio * rsi\_14 \\
alpha\_vwapdev\_over\_macd & alpha\_close\_vwap\_diff\_rank / (macd\_diff + 1e-6) \\
alpha\_volume\_times\_macd & Volume * macd\_diff \\
alpha\_macd\_times\_lowdev & macd\_diff * alpha\_close\_near\_low \\
alpha\_macd\_times\_volatility & macd\_diff * std(Close, 10) \\
alpha\_macd\_rank\_times\_rsi & rank(macd\_diff) * rsi\_14 \\
alpha\_macd\_diff\_sign & sign(macd\_diff) \\
alpha\_macd\_rsi\_ratio & macd\_diff / (rsi\_14 + 1e-6) \\
alpha\_boll\_bandwidth & boll\_upper - boll\_lower \\
alpha\_close\_boll\_mid\_ratio & (Close - boll\_mid) / (boll\_upper - boll\_lower + 1e-6) \\
alpha\_macd\_rank\_times\_price & rank(macd\_diff) * Close \\
alpha\_rsi\_rank\_times\_close & rank(rsi\_14) * Close \\
alpha\_volatility\_times\_rsi & std(Close, 10) * rsi\_14 \\
alpha\_rsi\_vs\_50 & rsi\_14 - 50 \\
alpha\_rsi\_bounce\_strength & I(rsi\_14↑ \& \textless30) * alpha\_rsi\_vs\_50 \\
alpha\_macd\_cross & I(macd\_diff \textgreater{} 0 \& macd\_diff.shift(1) $ \le $0) \\
alpha\_macd\_cross\_strength & alpha\_macd\_cross * macd\_diff \\
alpha\_rsi\_rise\_velocity & rsi\_14 - rsi\_14.shift(3) \\
alpha\_body\_times\_rsi & alpha\_kline\_body\_strength * rsi\_14 \\
alpha\_price\_vs\_boll & I(Close < boll\_lower) * Close \\
alpha\_boll\_rebound & I(Close yesterday \textless
 boll\_lower \& Close today \textgreater{} boll\_lower) \\
alpha\_rsi\_bounce\_rank & rank(alpha\_rsi\_bounce\_strength) \\
alpha\_macd\_cross\_rank & rank(alpha\_macd\_cross\_strength) \\
\bottomrule
\end{tabular}
\caption{Constructed 40 alpha factor formulas based on price-volume-behavior logic}
\end{table}

\end{document}